\documentclass{article}
\usepackage[utf8]{inputenc}
\usepackage{xcolor}
\usepackage{cite}
\usepackage[cmex10]{amsmath}
\usepackage{framed}
\usepackage{mathtools}
\usepackage{caption}
\usepackage{amsfonts}
\usepackage{amssymb}
\usepackage{array}
\usepackage{graphicx}
\usepackage{amsthm}
\usepackage{enumitem}

\newtheorem{theorem}{Theorem}
\newtheorem{assumption}{Assumption}
\newtheorem{corollary}{Corollary}

\newtheorem{lemma}{Lemma}
\newtheorem{definition}{Definition}

\newtheorem{remark}{Remark}
\newtheorem{conjecture}{Conjecture}
\newcommand{\RE}{\ensuremath{\mathrm{Re}}}

\newcommand{\field}[1]{\mathbb{#1}}
\newcommand{\R}{\field{R}}

\newcommand{\C}{\field{C}}

\newcommand{\Z}{\field{Z}}

\newcommand{\bmtx}{\begin{bmatrix}}
\newcommand{\emtx}{\end{bmatrix}}
\newcommand{\bsmtx}{\left[ \begin{smallmatrix}} 
\newcommand{\esmtx}{\end{smallmatrix} \right]}

\def\Real{\mathbb{R}}

\title{On the  Necessity and Sufficiency of Discrete-Time O'Shea-Zames-Falb Multipliers}

\author{ Lanlan Su \thanks{L. Su is with the School of Engineering, University of Leicester, Leicester, LE1 7RH, UK. (ls499@leicetser.ac.uk)} \and Peter Seiler \thanks{P. Seiler is with the Department of Electrical Engineering \& Computer Science, University of Michigan, Ann Arbor, US. (pseiler@umich.edu)} \and Joaquin Carrasco \thanks{ J. Carrasco is with the Department of Electrical \& Electronic Engineering, University of Manchester, M13 9PL, UK. (joaquin.carrasco@manchester.ac.uk) } \and Sei Zhen Khong \thanks{S. Khong is an independent researcher. (szkhongwork@gmail.com)}
}

\date{\vspace{-5ex}}
\begin{document}
\maketitle

\begin{abstract}
 This paper considers the robust stability of a discrete-time Lurye system consisting of the feedback interconnection between a linear system and a bounded and monotone nonlinearity. It has been conjectured that the existence of a suitable linear time-invariant (LTI) O'Shea-Zames-Falb multiplier is not only sufficient but also necessary. Roughly speaking, a successful proof of the conjecture would require: (a) a conic parameterization of a set of multipliers that describes exactly the set of nonlinearities, (b) a lossless S-procedure to show that the non-existence of a multiplier implies that the Lurye system is not uniformly robustly stable over the set of nonlinearities, and (c) the existence of a multiplier in the set of multipliers used in (a) implies the existence of an LTI multiplier. We investigate these three steps, showing the current bottlenecks for proving this conjecture. In addition, we provide an extension of the class of multipliers which may be used to disprove the conjecture.
 
\end{abstract}

\section{Introduction}

A class of noncausal linear time-invariant (LTI) multipliers preserving the positivity of monotone nonlinearities was proposed by O'Shea for continuous-time~\cite{o1967improved} and discrete-time~\cite{OShea67}. In continuous-time, Zames and Falb~\cite{Zames68} produced a rigorous treatment of noncausal LTI multipliers to show that the existence of a suitable multiplier is a sufficient condition for the stability of the Lurye system with monotone and bounded nonlinearities. In discrete-time, similar development was produced by Willems and Brockett~\cite{Willems68}, but they extended the original class by including linear time-varying (LTV) multipliers. Nowadays, both continuous-time and discrete-time  LTI classes of multipliers are referred to as O'Shea-Zames-Falb (OZF), or just Zames-Falb, multipliers.

Although the continuous-time class of OZF multipliers has attracted more attention (see~\cite{carrasco2016zames} for a recent overview), its discrete-time counterpart has attracted significant attention in the past years. Convex searches leading to numerical criteria have been proposed in~\cite{Fetzer17b,Carrasco20,Turner:21}, and it is worth highlighting its role in convergence analysis of optimisation algorithms, e.g.~\cite{Lessard16,Freeman18,Zhang20,Michalowsky21,lee2020}. The efficiency of the searches for discrete-time OZF multipliers has generated questions on the conservatism of the sufficient condition with OZF multipliers, e.g. investigations into phase limitations \cite{Megretski95,jonsson1996stability,wang2017phase} and limits derived from duality theory\cite{Zhang21}.
It was conjectured by Carrasco that the existence
of a suitable OZF multiplier is, in fact, both necessary and sufficient for robust stability~\cite{wang2017phase}.  The conjecture motivated the discovery of the first second-order counterexamples to the discrete-time Kalman conjecture~\cite{Heath15} and also  motivated  a systematic construction of destabilizing nonlinearities~\cite{Seiler20}.

This paper analyses the Carrasco conjecture. Firstly, we show that the class of LTV multipliers proposed in~\cite{Willems68} exactly characterises the class of monotone nonlinearities and it can be parameterized in terms of conic combinations. Secondly, we show a necessity condition for an extension of the class of nonlinearities. Thirdly, we show that the existence of a suitable LTV multiplier for an LTI system implies the existence of a suitable LTI multiplier, which belongs to the class of OZF multipliers. Our analysis identifies that the requirement of a countably infinite class of multipliers may imply an inherent conservatism, as the current version of the lossless S-procedure is limited to a finite number of constraints. The use of a finite, although arbitrarily large, number of constraints in the S-procedure becomes the main bottleneck to solving the Carrasco conjecture.  As a result, the conjecture remains unsolved. 
As a means to disprove the Carrasco conjecture, we introduce a set of nonlinear multipliers and show that it fully characterises monotone static nonlinearities. The Carrasco conjecture would thus be disproved if one could find \emph{an} LTI system that cannot be characterised by any OZF multiplier but can be characterised by a nonlinear multiplier we proposed. This significantly contrasts with an alternative to disproving the conjecture, which involves showing that there always exists a destabilizing nonlinearity for \emph{every} LTI system not characterisable by any OZF multiplier.

The structure of the paper is a follows. Section \ref{sec: pre} describes the notation and provides the preliminaries of the paper. Section \ref{sec: problem} provides a formal problem statement. Section \ref{sec: main} provides the main technical results of the paper. Initially, we identify a class of nonlinearities that includes all monotone nonlinearities, over which the robust stability of the Lurye system is ensured by the existence of a suitable finite-impulse response LTI multiplier. The same condition is shown to be necessary for the robust stability of the Lurye system when the set of nonlinearities is replaced by a relevant relation set. As an intermediate step, we use a wider class of LTV multipliers, and show the Lurye system with a periodic LTV plant is robustly stable if  there exists a suitable LTV multipliers. Then, for LTI plants, we show that the existence of a suitable LTV multiplier is necessary and sufficient for the existence of a suitable LTI multiplier. Moreover, the links between the identified set of nonlinearities  and the set of monotone nonlinearities are established through the sets of LTV multipliers that  characterise them, whereby the discrete-time  version of the classical Zames-Falb theorem is recovered. Finally, Section \ref{sec: extention} extends the class of multipliers by using nonlinear multipliers which may be useful to disprove the discrete-time Carrasco conjecture.

\section{Notation and Preliminaries} \label{sec: pre}

Let $\R$, $\C$, $\Z$, $\Z_{0}^{+}$, $\Z_{0}^{-}$ and $\Z^+$ denote the sets of real numbers, complex numbers, integers, non-negative integers,  non-positive integers and positive integers, respectively. We use  $\RE \left\lbrace \lambda\right\rbrace$ to denote the real part of a complex number $\lambda$.

Define $\ell_2$ as the set of (two-sided) discrete-time signals $u:\Z \to \R^n$ where
$\sum_{k\in\Z} u_k^T u_k <\infty$.
This forms an inner product space with $\langle u,w\rangle \vcentcolon= \sum_{k\in\mathbb{Z}}u_k^T w_k$ and
corresponding norm $\|u\|:=[\langle u,u \rangle]^{1/2}$. We will also use one-sided
signals $\ell_2^{0+} \vcentcolon= \left\{ f \in \ell_2 :  f_i=0, \forall i<0 \right\}$. Two important operations, defined
for any $\tau\in \Z$, are
the truncation $P_\tau:\ell_2 \to \ell_2$
and (rightward) shift $S_\tau: \ell_2 \to \ell_2$. The truncation operator is defined by
$(P_\tau u)_k:=u_k$ for $k \le \tau$ and $(P_\tau u)_k:=0$ for $k> \tau$.  The shift operation is defined 
by $(S_\tau u)_k:=u_{k-\tau}$. Further define the two-sided truncation operator $P_{-\tau,\tau}$ as $(P_{-\tau,\tau}u)_k:=u_k$ for $k=-\tau,\ldots,\tau$ and $(P_{-\tau,\tau}u)_k:=0$ otherwise. 
The  one-sided extended space
is $\ell_{2e}^{0+}  := \left\lbrace f: \mathbb{Z} \rightarrow \mathbb{R}^n : P_\tau f\in \ell_2^{0+} ,\forall \tau\in \mathbb{Z}_0^+ \right\rbrace$.

A system (also called operator) $H$ is modeled as an operator that
maps an input sequence $u$ to an
output $y:=H u$. 
$H$ is said to be linear if $H(u+v) = Hu+Hv$ and $H(cu)=c\, Hu$ for any $u,v\in \ell_2$ and $c\in \R$.

For a system $M:\ell_2\rightarrow\ell_2$, 
the induced-$\ell_2$ norm is
defined as:
\begin{align*}
    \|M\|:= \sup_{0\neq u \in \ell_2}
    \frac{\|Mu\|}{\|u\|}. 
\end{align*}
$M$ is said to be bounded if $\|M\|<\infty$.
 The set of all bounded, linear operators mapping from $\ell_2$ to $\ell_2$ is denoted as $\mathcal{L}(\ell_2,\ell_2)$. The system $M\in \mathcal{L}(\ell_2,\ell_2)$ can be represented by a doubly-infinite matrix
of real numbers  $\{m_{ij}\}_{i,j\in\Z}$ such that $y=Mu$ is defined by $y_i=\sum_{j\in\Z} m_{ij}u_j$ for  $i\in\Z$. This infinite sum exists for all $u\in \ell_2$ and the resulting sequence belongs to  $\ell_2$.  $M$ is said to be time-invariant if $M S_\tau = S_\tau M$, for all $\tau\in\Z$.  $M$ is time-invariant if and only if its matrix representation is Toeplitz, i.e. $m_{i+l,j+l} = m_{i,j}$ for all $i,j,l \in \Z$.
Time-invariance  of $M$ implies the matrix representation is uniquely defined by $\overline{m}_i := m_{l+i,l}$ and the response
$y=Mu$ is given by the convolution $y_k=\sum_{l\in\Z} \overline{m}_{k-l} u_l$.
The diagonals of the Toeplitz matrix $\{ \overline{m}_i \}_{i\in \Z}$ are the impulse response coefficients for the linear, time-invariant (LTI) system $M\in\mathcal{L}(\ell_2,\ell_2)$.
An important fact
is that LTI systems in $\mathcal{L}(\ell_2,\ell_2)$ have an equivalent transfer function representation.
Specifically, let $\mathbb{L}_\infty$ denote the set of complex functions $\hat{M}:\C\to\C$ that are essentially bounded on the unit circle.  If $M\in \mathcal{L}(\ell_2,\ell_2)$
is time-invariant then it has a transfer
function $\hat{M} \in \mathbb{L}_\infty$
such that $y=Mu$ is equivalent to
multiplication in the frequency domain:
$\hat{y}(\omega) =\hat{M}(e^{j\omega}) \hat{u}(\omega)$ where $\hat{u}(\omega) := \sum_{n\in \Z} u_n e^{-j \omega n}$ and $\hat{y}(\omega) := \sum_{n\in \Z} y_n e^{-j \omega n}$.  For $T\in\mathbb{Z}^+$, $M$ is said to be $T$-periodic if its matrix representation satisfies $m_{i+T,j+T} = m_{i,j}$ for all $i,j \in \Z$. We use the notation $M^*$ to denote the adjoint system of $M$.

A system $G:\ell_{2e}^{0+}\rightarrow\ell_{2e}^{0+}$ is said to be causal if $P_\tau G P_\tau =P_\tau G$ for all $\tau\in\Z_0^{+}$. A system  $G:\ell_{2e}^{0+}\rightarrow\ell_{2e}^{0+}$ is said to be $T_0$-periodic if $GS_{\tau T_0}=S_{\tau T_0} G$ for all $\tau\in\Z_0^{+}$ and  is said to be time-invariant if it is periodic with $T_0=1$. A causal system $G:\ell_{2e}^{0+}\rightarrow\ell_{2e}^{0+}$ is said to be bounded if 
\begin{align*}
    \|G\|:= \sup_{ \tau \in\Z^+;0 \neq P_\tau u \in \ell_2^{0+}}
    \frac{\|P_\tau Gu\|}{\|P_\tau u\|}=\sup_{0\neq u \in \ell_2^{0+}}
    \frac{\|Gu\|}{\|u\|} <\infty. 
\end{align*}
The set of all bounded, linear operators from $\ell_{2e}^{0+}$ to $\ell_{2e}^{0+}$ is denoted $\mathcal{L}(\ell_{2e}^{0+},\ell_{2e}^{0+})$.  
Let  $\mathbb{H}_\infty$ denote subspace of $\mathbb{L}_\infty$ that is analytic in the unit disk.   If $G\in \mathcal{L}(\ell_{2e}^{0+},\ell_{2e}^{0+})$ is both time-invariant and causal then it has a transfer function $\hat{G}\in \mathbb{H}_\infty$. See \cite{dahleh1994control} for more details of linear discrete-time operators.

A nonlinearity $\phi: \ell_{2e}^{0+}  \to \ell_{2e}^{0+} $ is memoryless if there exists $N:\mathbb{R} \to \mathbb{R}$ such that $(\phi (v))_i=N(v_i)$ for all $i\in\mathbb{Z}_0^+$. The memoryless nonlinearity $\phi$ is bounded if  there exists a constant $C>0$ such  that $ |N(x)|\le C|x|$ $\forall x\in\R$. Note that boundedness implies that $N(0)=0$.
The memoryless nonlinearity $\phi$
is monotone if $x_1\ge x_2$ implies $N(x_1)\ge N(x_2)$.

\section{Problem Formulation} \label{sec: problem}
\label{sec:probform}

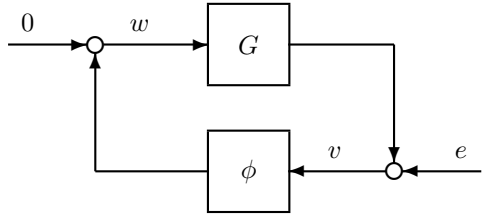
\begin{figure}[ht]
	\centering
\begin{picture}(180,75)(0,-65)
\thicklines
\put(0,0){\vector(1,0){30}}
\put(33,0){\circle{6}}
\put(5,5){0}

\put(36,0){\vector(1,0){40}}
\put(46,5){$w$}
\put(76,-15){\framebox(30,30){$G$}}

\put(106,0){\line(1,0){40}}
\put(146,0){\vector(0,-1){45}}
\put(146,-48){\circle{6}}

\put(179,-48){\vector(-1,0){30}}
\put(169,-43){$e$}
\put(143,-48){\vector(-1,0){37}}
\put(121,-43){$v$}
\put(76,-63){\framebox(30,30){$\phi$}}

\put(76,-48){\line(-1,0){43}}
\put(33,-48){\vector(0,1){45}}
\end{picture}

\caption{Lurye system}
\label{fig:luryesys}
\end{figure}

Let  $G\in \mathcal{L}(\ell_{2e}^{0+}, \ell_{2e}^{0+})$ model a discrete-time system that is single-input single-output, LTI, causal and bounded.
We consider the Lurye system of $G$ in 
feedback with a causal bounded nonlinearity $\phi: \ell_{2e}^{0+} \to \ell_{2e}^{0+}$ as shown Figure~\ref{fig:luryesys}. The Lurye system
$[G,\phi]$ is defined as:
\begin{align}
\label{eq:sys}
\begin{split}
    v &= Gw+e\\
    w &= \phi v.
\end{split} 
\end{align}
Note that the external signal at the input of $G$ has been set to zero.  This is done  without loss of generality as the effect of a non-zero external signal at the input of $G$ can be lumped with $e$ due to the assumption that $G$ is linear, causal and bounded.

 Well-posedness
and stability of the Lurye system
are defined next.

\begin{definition} \label{def:wellposedness}
  $[G,\phi]$ is \underline{well-posed} if  for any $e\in \ell_{2e}^{0+}$ there exist $v,w\in \ell_{2e}^{0+}$ that satisfy \eqref{eq:sys} and depend causally on $e$. 
\end{definition}
  
\begin{definition}\label{def:stability}
$[G,\phi]$ is \underline{stable} if it is well-posed and there exists $\gamma > 0$ such that 
\[
\sup_{\tau\in\Z^+,\; 0\neq P_\tau e\in \ell_2^{0+}} \frac{\|P_\tau w\|}{\|P_\tau e\|} \le \gamma .
\] 
\end{definition}

Let $\mathcal{S}_0$ be a set of nonlinearities that map 0 to 0.

\begin{definition}\label{def:uniform stability}
 $[G,\phi]$ is
  \underline{uniformly robustly stable
  over $\mathcal{S}_0$}
  if $[G,\phi]$ is well-posed for all
  $\phi \in \mathcal{S}_0$ and there exists
  $\gamma>0$ such that 
\begin{align*}
\sup_{\phi\in\mathcal{S}_0} \;
\sup_{\tau\in\Z^+,\; 0\neq P_\tau e\in \ell_2^{0+}}
\frac{\|P_\tau w\|}{\|P_\tau e\|}\le \gamma .
\end{align*}
\end{definition}

Two classes of nonlinearities will be consider in this work. In particular, the set of all nonlinearities  that are  memoryless, bounded, monotone is denoted $\mathcal{S}$. 
In Section \ref{subsec:larger set} we will introduce a second related set of nonlinearities denoted $\mathcal{S}^{T,B}$.

There is a large literature on robust stability of Lurye systems and
details can be found in \cite{D&V:75,willems1970analysis}.
The plant $G\in \mathcal{L}(\ell_{2e}^{0+},\ell_{2e}^{0+})$ is time-invariant and causal. Thus it has an equivalent transfer function representation $\hat{G}(e^{j\omega})\in \mathbb{H}_{\infty}$.  For continuous-time systems, it is shown in \cite{KHONG:21} that uniform robust stability over the set of monotone nonlinearities is guaranteed by the existence of an LTI Zames-Falb multiplier satisfying a certain frequency domain inequality. In what follows, two sets of multipliers for the discrete-time counterpart are introduced.

\begin{definition}\label{def:LTV}
  The class of LTV multipliers $\mathcal{M}_{\text{LTV}}\subset \mathcal{L}(\ell_2,\ell_2)$ is given by the set of linear operators satisfying the following condition:
  \begin{enumerate}
      \item The associated matrix $M=[m]_{ij}$ is a doubly hyperdominant matrix with zero excess, i.e., $m_{ij}\le 0, \forall i\neq j$ and $\sum_{i\in\mathbb{Z}}m_{ij}= 0, \forall j\in\mathbb{Z}$, $\sum_{j\in\mathbb{Z}}m_{ij}= 0, \forall i\in \mathbb{Z}$.
      \item For $\epsilon>0$ there exists $n=n(\epsilon)$ such that in  each row or each column the sum of $n$ entries with  largest absolute values is at most $\epsilon$.
  \end{enumerate} 
\end{definition}
It is known that the  finite-dimensional doubly hyperdominant matrices with zero excess   are precisely the convex combinations of  permutation matrices subtracted from $I$ of the same dimension  \cite[Theorem 3.7]{willems1970analysis}. To derive an infinite-dimensional version of this fact,  the second condition in Definition \ref{def:LTV} is needed. This will be elaborated in the proof of Lemma \ref{le:basis_infinite} in the next section.
 The class of LTI multipliers $\mathcal{M}_{\text{LTI}}\subset \mathcal{L}(\ell_2,\ell_2)$ is defined as 
 \begin{align}
       \mathcal{M}_{LTI} := \{M\in\mathcal{M}_{LTV}: M \text{ is LTI}\}.
 \end{align}

\begin{conjecture}\label{th:ZF}
Let  $G\in \mathcal{L}(\ell_{2e}^{0+}, \ell_{2e}^{0+})$ be LTI, causal and bounded.
Assume the Lurye system $[G,\phi]$ is well-posed for all $\phi \in \mathcal{S}$.
The feedback interconnection $[G,\phi]$ is uniformly robustly stable over $\mathcal{S}$  if and only if there exists  $M\in \mathcal{M}_{LTI}$  such that
\begin{equation}\label{eq: LTI_frequncyCon0}
  \RE \left\lbrace  \hat{M}(e^{j\omega}) G(e^{j\omega})\right\rbrace < 0,\forall \omega\in [0,2\pi].  
\end{equation}
\end{conjecture}

The sufficiency of Conjecture 1  is given in~\cite{willems1970analysis}. The main results of this paper analyse the necessity direction in Conjecture 1, also known as the Carrasco Conjecture.

\section{Main Results} \label{sec: main}

The main results are given in four different subsections. Firstly, we show that an exact characterisation of monotone nonlinearities involving conic parameterization. Secondly, we define a larger set of nonlinearities which will be used in the necessary results. We show a limiting argument that relate this set with the set of memoryless nonlinearities. Thirdly, we provide necessary conditions involving a finite-dimensional set of multipliers. Finally, we show the equivalence between LTV and LTI multipliers.

\subsection{Characterization of Monotone Nonlinearities}
\label{sec:charnl}

In this subsection, we show that the set of
input-output pairs of all nonlinearities in  $\mathcal{S}$ can be fully characterised by the set of time-varying multipliers $\mathcal{M}_{LTV}$. The presentation draws from \cite{Willems68} and Section 3.5 in  \cite{willems1970analysis}.  To start, consider finite sequences of real numbers $\{v_1,v_2,\ldots,v_n\}$ and $\{w_1,w_2,\ldots, w_n\}$. The sequences are similarly ordered if $v_i<v_j$ implies that $w_i\le w_j$. The sequences are unbiased if $v_iw_i\ge 0$ for $1\le i\le n$.  The definition of similarly ordered, unbiased $\ell_2$ sequences  is analogous to those for finite sequences. A finite-dimensional matrix $M\in\mathbb{R}^{n\times n}$ is said to be doubly hyperdominant if $m_{ij}\le 0$ for $i\neq j$ and $\sum_{i=1}^n m_{ij}\ge 0$ and $\sum_{j=1}^{n}m_{ij}\ge 0$ for all $i,j\in \{1,\ldots,n\}$. The definition of doubly hyperdominance for doubly-infinite matrices is analogous to that for finite-dimensional matrices. 

Our results will relate the following three sets of sequences:
\begin{align*}
\mathcal{G}_1 & :=\{ (v, w) \in \ell_2 : w = \phi v, \, \phi \in \mathcal{S} \} \\
\mathcal{G}_2 & := \{ (v, w) \in \ell_2 : (v, w) \text{ is similarly ordered}\} \\
\mathcal{G}_3 & := \{ (v, w) \in \ell_2 : 
\langle v, Mw \rangle \ge 0, \,
\forall M \in \mathcal{M}_{LTV} 
\} 
\end{align*}
A fourth useful set of sequences will also be introduced later in this subsection. It is worth mentioning that two sequences in $\ell_2$ are similarly ordered if and only if they are similarly ordered and unbiased \cite[P63]{willems1970analysis}. 
Therefore,  $\mathcal{G}_2$ can be equivalently expressed as 
\[\mathcal{G}_2= \{ (v, w) \in \ell_2 : (v, w) \text{ is similarly ordered and unbiased}\}.\]
The first lemma relates the input-output pairs of nonlinearities in $\mathcal{S}$ and  similarly ordered sequences.

\begin{lemma} \label{le: IO equivalence}
The set $\mathcal{G}_1  :=\{ (v, w) \in \ell_2 : w = \phi v, \, \phi \in \mathcal{S} \}$ is equal
to the closure of the set
$\mathcal{G}_2 := \{ (v, w) \in \ell_2 : (v, w) \text{ is similarly ordered}\}$.
\end{lemma}
\noindent\textbf{Proof.} ($\subset$) This follows from the fact that any $(v, w)\in \ell_2$ satisfying $w = \phi v$ for a $\phi \in \mathcal{S}$ is necessarily similarly ordered and unbiased.

($\supset$) Let $(v, w) \in \ell_2$ be
similarly ordered, then $(v, w)$ is unbiased. Given any $\epsilon > 0$, there exists a finite $\tau>0$ such that
$(\bar{v},\bar{w}):=(P_{-\tau,\tau} v,P_{-\tau,\tau}w)$ satisfy
$\|v-\bar{v}\| \le \epsilon/2$, and $\|w-\bar{w}\| \le \epsilon/2$. The truncated sequences have a finite number of unique pairs coming from  $(\bar{v}_k,\bar{w}_k)_{k=-\tau}^\tau$ and $(\bar{v}_k,\bar{w}_k)=(0,0) $ otherwise. The truncated sequences are similarly ordered and unbiased so the data can be linearly interpolated by a monotone function.  However, the function could be multi-valued and/or unbounded if the data contains points with $\bar{v}_i=\bar{v}_j$ but $\bar{w}_i \ne \bar{w}_j$.
This issue is resolved by another perturbation to the data. Specifically, define $\hat{w}:=\bar{w}$ and define $\hat{v}$ by perturbing $\bar{v}$ by a sufficiently small $\delta>0$ as follows.  
If $\bar{v}_i=0$ but $\bar{w}_i \ne 0$ for any $i\in \Z$ then define $\hat{v}_i := \delta \bar{w}_i$. This preserves the point $(0,0)$ and perturbs other pairs to lie along a line of slope $\delta$.
Similarly, suppose the sequence $\bar{v}$ has a non-zero value repeated $N$ times: $\bar{v}_{i_1}=\cdots=\bar{v}_{i_N}\ne 0$
with $\bar{w}_{i_1} \le  \cdots
\le \bar{w}_{i_N}$ for some indices $\mathcal{I}:=\{i_1,\ldots,i_N\}$. In this case, define $\hat{v}_k=\bar{v}_{i_1} + \delta (\bar{w}_k - \bar{w}_{i_1})$ for $k\in \mathcal{I}$. Again, this perturbs the repeated pairs to lie along a line of slope $\delta$. The perturbed sequences $(\hat{v},\hat{w})$ can be linearly interpolated by a single-valued, monotone, bounded function $N : \mathbb{R} \to \mathbb{R}$ such that $N(0) = 0$ and $N(\hat{v}_i) = \hat{w}_i$ for all $i \in \mathbb{Z}$. Moreover, we have $\|\bar{w}-\hat{w}\| =0$ by definition and, 
if $\delta>0$ is sufficiently small, then 
$\|\bar{v}-\hat{v}\| \le \epsilon/2$.

In summary, for any $\epsilon > 0$ and similarly ordered, unbiased $(v, w) \in \ell_2$, there exist a
pair $(\hat{v},\hat{w}) \in \ell_2$ such that: (i) $\|v-\hat{v}\| \le \epsilon$,  (ii) $\|w-\hat{w}\| \le \epsilon$, and (iii) $\hat{w}=\phi(\hat{v})$ for some $\phi\in \mathcal{S}$. The inclusion holds since $\epsilon$ can be arbitrarily small.
\hfill\qed

The next lemma states that similarly ordered sequences are exactly characterised by a positivity condition involving $\mathcal{M}_{LTV}$.

\begin{lemma}\label{le: TV_IQC_similar sequences}
The set $\mathcal{G}_2 := \{ (v, w) \in \ell_2 : (v, w) \text{ is similarly ordered}\}$
is equal to the set
$\mathcal{G}_3 := \{ (v, w) \in \ell_2 : 
\langle Mv, w \rangle \ge 0, \,
\forall M \in \mathcal{M}_{LTV}\}$.
\end{lemma}
\noindent \textbf{Proof.}
($\subset$)  This follows from Theorem 3.11  \cite{willems1970analysis} as $\mathcal{M}_{LTV}$ is a subset of the set consisting of all $M\in\mathcal{L}(\ell_2,\ell_2)$ whose associated matrix is doubly hyperdominant. 

($\supset$) Assume
$\langle Mv,w \rangle \ge 0$ for every $M\in \mathcal{M}_{LTV}$. First consider the multiplier $M$
with the associated matrix defined by:
\begin{align*}
\begin{bmatrix} m_{kk} & m_{kl}\\
m_{lk} & m_{ll}
\end{bmatrix} =  \begin{bmatrix}1 & -1\\
-1 & 1
\end{bmatrix}
\mbox{ and }
m_{ij}  =  0, \text{otherwise}.
\end{align*}
By construction $M\in \mathcal{M}_{LTV}$. Moreover,
$\langle Mv,w \rangle \ge 0$ can be rewritten as $(v_k-v_l)(w_k-w_l) \ge 0$. Thus
if $v_k<v_l$ then $w_k\le w_l$ and
the sequences $v,w$ are similarly ordered.
 \hfill\qed

\vspace{2mm}

Next, we introduce one final representation for the sets of sequences related to $\phi\in \mathcal{S}$. Recall that a matrix $M$ is said to be doubly stochastic if all entries are non-negative and all rows and columns sum to $1$. Let $\mathcal{P}$ denote the set of operators in $\mathcal{L}(\ell_2,\ell_2)$ whose matrix representation is a doubly-infinite  permutation matrix.  Define 
\begin{align}\label{eq: basis_inf}
   \mathcal{C}:=\{C \in \mathcal{L}(\ell_2,\ell_2): C=I-P,P\in \mathcal{P}\}.  
\end{align}
Observe that each element in $\mathcal{C}$ has a matrix representation which is doubly hyperdominant with zero excess, i.e., the sum of each row and each column is zero.  

The next  lemma provides a useful representation for the set $\mathcal{M}_{LTV}$.
\begin{lemma}\label{le:basis_infinite}
The set of all conic combinations of $\mathcal{C}$ is equal to $\mathcal{M}_{LTV}$.
\end{lemma}
\noindent\textbf{Proof.}  
First, we introduce two statements \ref{d1} and \ref{d2} that are used in the proof below.
\begin{enumerate}[label=(C\arabic*)]
\item\label{d1} for every $\delta>0$ there exists  $n=n(\delta)$ such that in each row or column the sum of the $n$ largest entries is at least $1-\delta$.  
\item\label{d2} for every $\epsilon>0$ there exists $n=n(\epsilon)$ such that in each row or column the sum of the $n$ entries with  largest absolute values is at most $\epsilon$.  
\end{enumerate} 

Next, define  
\[\mathcal{A} := \{ A \in\mathcal{L}(\ell_2,\ell_2): A \text{ is doubly-infinite, doubly stochastic and satisfies (C1)}\}.
\]
In what follows we show that 
\begin{align}\label{eq: AD}
   \mathcal{M}_{LTV} = \{d(I-A)\in \mathcal{L}(\ell_2,\ell_2) : d>0, A \in \mathcal{A}\}. 
\end{align}
To show ($\subset$), let $D$ be any  doubly-infinite  doubly hyperdominant matrix with zero excess, and define $d:=\max_{i,j} |d_{ij}|$. Then  $D$  can be expressed as $D=d(I-A)$ where $A:=\frac{1}{d} (dI-D)$ is a doubly-infinite   doubly stochastic matrix.  If $D$ satisfies \ref{d2}  then $A$ satisfies \ref{d1}. To show ($\supset$),   let $d$ be any positive real number and $A$ be any element in $\mathcal{A}$. It is clear that $d(I-A)$  is  doubly hyperdominant  with zero excess and satisfies  \ref{d2}. 

Note that $\mathcal{A}$ is equal to the convex closure of $\mathcal{P}$ (\cite{isbell1955}).  This fact, together with \eqref{eq: AD}, implies that $\mathcal{M}_{LTV}$ is equal to conic combinations of elements in $\mathcal{C}$.

\hfill\qed

\vspace{2mm}

Define $\mathcal{G}_4 := \{ (v, w) \in \ell_2 : 
\langle Mv, w \rangle \ge 0, \,
\forall M \in \mathcal{C}\}$.   Combining Lemmas \ref{le: IO equivalence}, \ref{le: TV_IQC_similar sequences} and \ref{le:basis_infinite} yields the following theorem. 

\begin{theorem}\label{th:super}
Consider the following sets of sequences:
\begin{align*}
\mathcal{G}_1 & :=\{ (v, w) \in \ell_2 : w = \phi v, \phi \in \mathcal{S} \} \\
\mathcal{G}_3 & := \{ (v, w) \in \ell_2 : 
\langle Mv, w \rangle \ge 0, \,
\forall M \in \mathcal{M}_{LTV} \}\\
\mathcal{G}_4& := \{ (v, w) \in \ell_2 : 
\langle Mv, w \rangle \ge 0, \,
\forall M \in \mathcal{C} 
\} 
\end{align*}
$\mathcal{G}_1$ is equal to the closure of $\mathcal{G}_3$. $\mathcal{G}_1$ is also equal to the closure of $\mathcal{G}_4$.  
\end{theorem}

\noindent \textbf{Proof.} 
That $\mathcal{G}_1$ is equal to the closure of $\mathcal{G}_3$ can be obtained by combining Lemmas \ref{le: IO equivalence} and \ref{le: TV_IQC_similar sequences}. To prove that $\mathcal{G}_1$ is equal to the closure of $\mathcal{G}_4$, it suffices to show $\mathcal{G}_3=\mathcal{G}_4$. If $(v,w)$ is in $\mathcal{G}_3$, then $\langle Mv, w \rangle \ge 0$ for all $M \in \mathcal{M}_{LTV}$. Since $\mathcal{C}\in\mathcal{M}_{LTV}$, this implies  that $(v,w)$ is in $\mathcal{G}_4$. Thus, $\mathcal{G}_3\subset\mathcal{G}_4$. If $(v,w)$ is not in $\mathcal{G}_3$, then $\langle Mv, w \rangle < 0$ for some $M \in \mathcal{M}_{LTV}$. We can express such $M$ as a conic combination of elements $M_i\in\mathcal{C}$ according to Lemma \ref{le:basis_infinite}. Then, it follows that $\langle M_i v,w \rangle <0$ for at least one $M_i\in\mathcal{C}$. Hence, $(v,w)$ is not in $\mathcal{G}_4$. By contraposition $\mathcal{G}_4\subset\mathcal{G}_3$.  
\hfill\qed

\vspace{2mm}

Note that the set $\mathcal{C}$ contains a countable number of elements, and by Theorem~\ref{th:super} the set $\mathcal{S}$ is completely captured by the class of  multipliers defined by $\mathcal{C}$. 

It should be remarked that Theorem~\ref{th:super}  and the preceding lemmas  remain true with $v,w\in \ell_2^{0+}$ since $\ell_2^{0+}\subset\ell_2$.

\subsection{A Larger Set of Nonlinearities}\label{subsec:larger set}

In this subsection, we introduce a larger set of nonlinearities which contains all  nonlinearities in $\mathcal{S}$ as a subset. This  set facilitates the proof of the necessity of the existence of a certain multiplier for robust stability.

To this end, let $\pi(\cdot):\mathbb{Z}\rightarrow\mathbb{Z}$ denote a permutation. For $T,B \in\mathbb{Z}^+$,  define $\mathcal{P}_{T,B}$ as the set of $\pi(\cdot)$ that satisfies  $\pi(k+T)=\pi(k)+T,\forall k\in\mathbb{Z}$ and $\pi(i)\neq j$, for all $i,j\in\mathbb{Z}$ such that $|i-j|> B$.  Given any $T\in\Z^+$, if $\pi(k+T)=\pi(k)+T,\forall k\in\mathbb{Z}$, then evidently it holds that $\pi(k+nT)=\pi(k)+nT,\forall k\in\mathbb{Z},n\in\Z^+$. Thus, by replacing $T$ with $nT$ for some large enough $n\in\Z^+$,   we can assume without loss of generality that $T\ge 2B+1$. 

Further define the set of sequences:
\begin{align}
    \mathcal{G}^{T,B}:=\{(v,w)\in \ell_2: \sum_{k\in\mathbb{Z}}v_kw_k\ge \sum_{k\in\mathbb{Z}}v_{\pi(k)}w_k, \forall \pi\in \mathcal{P}_{T,B}\}
,\end{align}
and
\begin{align}
\mathcal{G}_1^{T,B}:=\text{the closure of }\mathcal{G}^{T,B}.
\end{align}

In what follows, we define a set of multipliers that are related to  $\mathcal{G}^{T,B}$. 
Given $T\in\mathbb{Z}^+$, we define the set of all $T$-periodic elements in $\mathcal{M}_{LTV}$:
\begin{align*}
    \mathcal{M}_{LTV}^{T}:=\{M\in\mathcal{M}_{LTV}:  M \text{ is $T$-periodic}\}.
\end{align*}
Given $B\in\mathbb{Z}^+$, further define 
\begin{align*}
    \mathcal{M}_{LTV}^{T,B}:=\{M\in\mathcal{M}_{LTV}^{T}:  m_{ij}=0, \; \forall  |i-j|> B\}.
\end{align*}
Note that $ \mathcal{M}_{LTV}^{T,B}$ consists of  all ``banded'' elements with bandwidth $B$ in $\mathcal{M}_{LTV}^{T}$. Recall from Lemma \ref{le:basis_infinite} that $\mathcal{M}_{LTV}$ is equal to the set of all conic combinations of $\mathcal{C}$ where $\mathcal{C}$ is defined in \eqref{eq: basis_inf}. To derive an analogous result for the case with $\mathcal{M}_{LTV}^{T,B}$, define $\mathcal{C}^{T,B}:=\{C\in\mathcal{C}:  C \text{ is $T$-periodic},\;  c_{i,j}=0, \forall |i-j| > B\}$. 

\begin{lemma}\label{le: basis_TB}
Let $T,B$ in $\Z^+$ be given with $T\ge2B+1$. The set $\mathcal{M}_{LTV}^{T,B}$ is equal to the set of all conic combinations of $\mathcal{C}^{T,B}$. 
\end{lemma}
\noindent\textbf{Proof.} 
($\supset$) This follows from Lemma \ref{le:basis_infinite} and the fact that the sum of $T$-periodic $B$-banded matrices is still $T$-periodic  and $B$-banded.

($\subset$)  For ease of exposition, we  prove in the following for the case with $T=4 $ and $B=1$. Note that the same reasoning lines can be applied to prove for the general cases with any $T,B\in\Z^+$ satisfying $T\ge 2B+1$.  
\begin{figure}
\begin{center}
\includegraphics[height=6cm]{{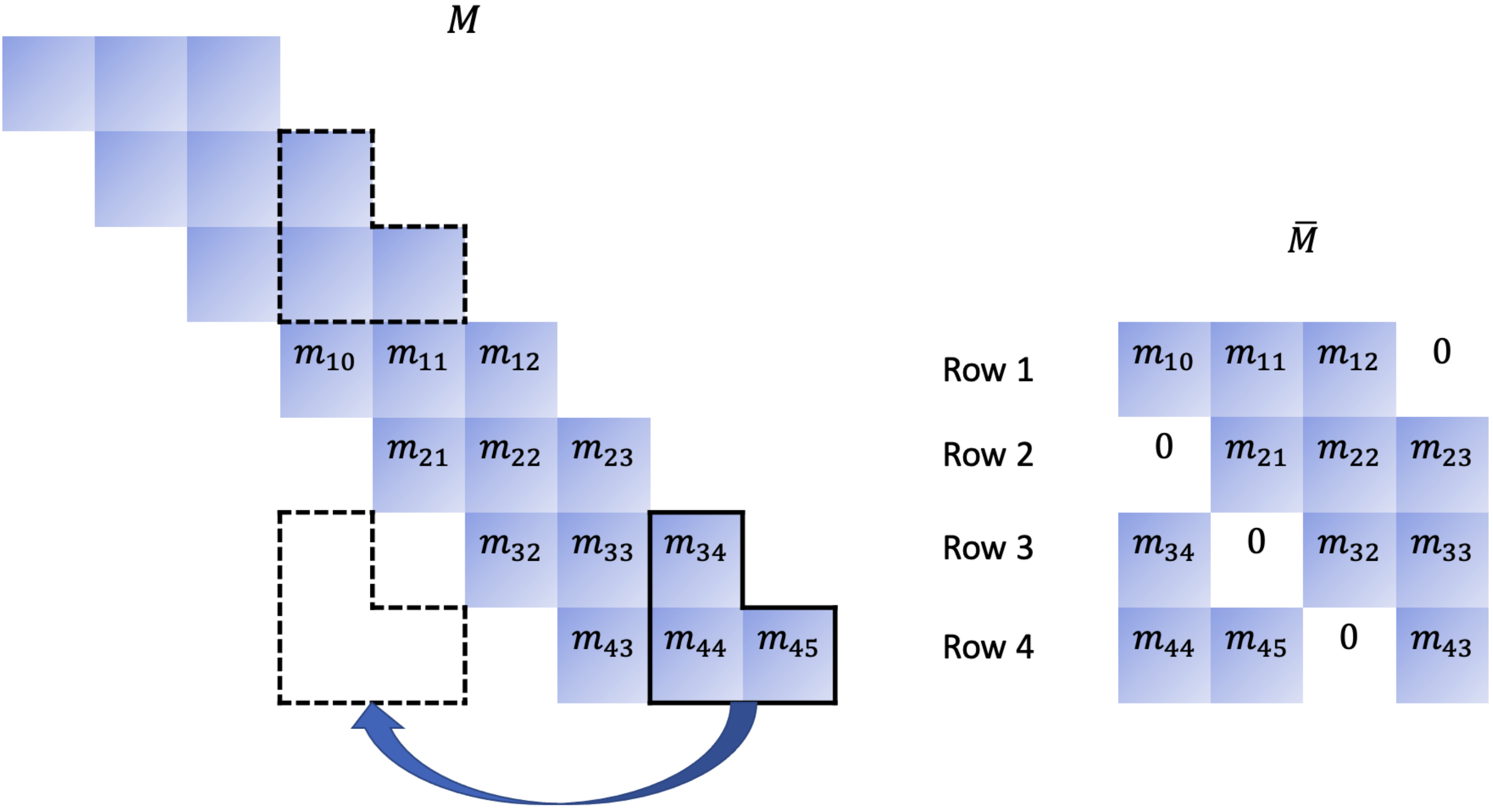}}    % The printed column  
 % width is 8.4 cm.
\end{center} 
 \caption{Left: a representation of $M$ and the translation. Right: $\bar{M}.$  }     
 \label{fig:proof_1}                            % accordingly.
\end{figure}Any $M \in \mathcal{M}_{LTV}^{4,1}$ can be described by the banded matrix in Figure \ref{fig:proof_1} (Left). Therein, the  shaded entries are possibly nonzero while the other entries are all zeros.  Since $M$ is $4$-periodic, the associated matrix of $M$ can be represented by any $4$ consecutive rows, say Row $1$ to Row $4$.  By definition, all rows and columns of $M$ sum to zero. Hence, the sum of Rows $1,2,3,4$ and Columns $0,1,2,3$ are zeros. Now  we translate horizontally the entries $m_{34},m_{44},m_{45}$ to the left by four steps, as indicated in Figure \ref{fig:proof_1} (Left). Then apparently the row sum for Rows $1,2,3,4$  remain the same. Moreover, the fact that $M$ is $4$-periodic implies that the translated entries inside the dashed border  in Rows $3,4$ can be equivalently obtained by translating the entries  inside the dashed border in Rows $-1,0$ vertically downward by four steps. Therefore,  the $4\times4$ matrix  $\bar{M}$ as shown in  Figure \ref{fig:proof_1} (Right), which is obtained by by extracting Rows $1,2,3,4$ and Columns $0,1,2,3$ of  Figure \ref{fig:proof_1} (Left) after translation,  is doubly hyperdominant with zero excess. Next, let $\mathcal{P}_4$ denote the set of $4\times4$ permutation matrices. By the arguments of Lemma \ref{le:basis_infinite} tailored to finite-dimensional matrices, we have that $\bar{M}$ can be expressed as a conic combination of $\mathcal{C}_4:=\{\bar{C}\in\R^{4\times4}: \bar{C}= I_4-P,P\in \mathcal{P}_4\}$. In other words,   there exist $\alpha_i> 0,i=1,\ldots,n$ with $n\le 4!$ such that $\bar{M}=\sum_{i=1}^{n} \alpha_i \bar{C}_i$ with $\bar{C}_i\in \mathcal{C}_4$. Note that every off-diagonal entry of all $\bar{C}\in \mathcal{C}_4$ takes values in $\{-1,0\}$. Hence, if the ${ij}$-th entry in $\bar{M}$ is zero, then  the ${ij}$-th entry in all $\bar{C}\in \{\bar{C}_1,\ldots,\bar{C}_n\}$ must be zero. Let $\bar{C}_i$  be any element in $\{\bar{C}_1,\ldots,\bar{C}_n\}$, depicted in Figure \ref{fig:proof_1} (Left).
\begin{figure}
\begin{center}
\includegraphics[height=6cm]{{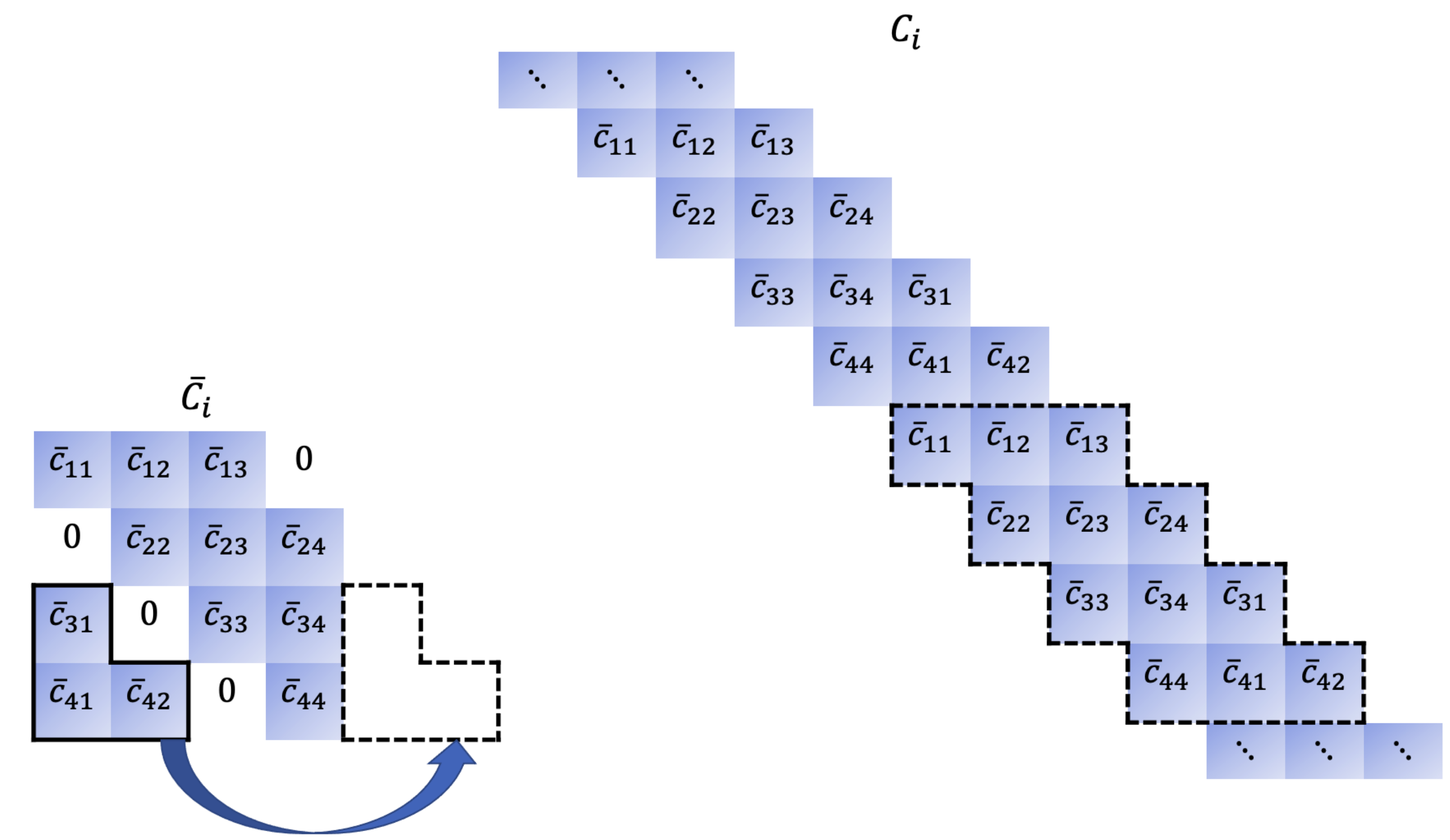}}    % The printed column  
 % width is 8.4 cm.
\end{center} 
 \caption{Left: $\bar{C}_i$ and the translation. Right: $C_i$. }     
 \label{fig:proof_2} 
 \end{figure}
 By reversing the process in Figure \ref{fig:proof_1}, we translate horizontally the entries $\bar{c}_{31},\bar{c}_{41},\bar{c}_{42}$ in $\bar{C}_i$ to the right by four steps, as shown in  Figure \ref{fig:proof_2} (Left), whereby  a banded array is obtained. Next, by repeating the obtained banded array  as shown in  Figure \ref{fig:proof_2} (Right), we get a banded and periodic matrix $C_i$. Since $\bar{M}=\sum_{i=1}^{n} \alpha_i \bar{C}_i$, it follows from the operations described in Figures \ref{fig:proof_1} and \ref{fig:proof_2}  that $M=\sum_{i=1}^{n}\alpha_iC_i$. It can also be observed that $C_i$ is a permutation matrix that is banded and $4$-periodic, i.e., $C_i\in\mathcal{C}^{4,1}$. Hence, each element in $\mathcal{M}^{4,1}$ can be expressed as a conic combination of $\mathcal{C}^{4,1}$. 
\hfill\qed

The next theorem  states that $\mathcal{G}^{T,B}$ can be exactly characterised by a positivity condition involving $ \mathcal{M}_{LTV}^{T,B}$ or $\mathcal{C}^{T,B}$.

\begin{theorem}\label{th:GTB_pos}
Let $T,B$ in $\Z^+$ be given with $T\ge2B+1$. The set $\mathcal{G}^{T,B}$  is equal to $\{(v,w)\in \ell_2:\langle Mv,w\rangle\ge 0, \forall M\in \mathcal{M}_{LTV}^{T,B}\}$. The set $\mathcal{G}^{T,B}$ is also equal to  $\{(v,w)\in \ell_2:\langle Mv,w\rangle\ge 0, \forall M\in \mathcal{C}^{T,B}\}$.
\end{theorem}
\noindent\textbf{Proof.} By definition, $\mathcal{G}^{T,B}$ can be expressed as
\[
\mathcal{G}^{T,B} = \{(v,w)\in \ell_2: \langle v,w\rangle \ge \langle  P v,w \rangle, \forall P\in \mathcal{P}^{T,B}\},
\]
where $\mathcal{P}^{T,B}$ is the set of all operators $P$  whose matrix representations are permutation matrices satisfying that  $p_{ij}=p_{i+T,j+T},\forall i,j\in\mathbb{Z}$ and $p_{ij}=0,\forall |i-j|>B$.  By definition, we have that the set $\{I-P: P\in \mathcal{P}^{T,B}\}$ is equal to $\mathcal{C}^{T,B}$. Therefore, we have
\begin{align*}
  \mathcal{G}^{T,B} =\{(v,w)\in \ell_2:\langle Mv,w\rangle\ge 0, \forall M\in \mathcal{C}^{T,B}\}.  
\end{align*}
Then, it follows from Lemma \ref{le: basis_TB}  that the set of all conic combinations of $\mathcal{C}^{T,B}$ is equal to $\mathcal{M}_{LTV}^{T,B}$. This, in turn, implies that  
\begin{align*}
  \mathcal{G}^{T,B} =\{(v,w)\in \ell_2:\langle Mv,w\rangle\ge 0, \forall M\in \mathcal{M}_{LTV}^{T,B}\}.  
\end{align*}
\hfill\qed

Since every $C\in\mathcal{C}^{T,B}$ has a finite bandwidth $B$ and a finite period $T$, the number of  elements in $\mathcal{C}^{T,B}$, denoted as $1+N_{T,B}$,  is finite. Denote
\begin{equation}\label{eq:finite basis}
    \mathcal{C}^{T,B}=0\cup \{C_1,C_2,\ldots,C_{N_{T,B}}\}.
\end{equation}

For any $T,B\in\Z^+$, let $\mathcal{S}^{T,B}$ denote the set of all causal bounded nonlinearities $\phi:\ell_2\rightarrow\ell_2$ such that 
\begin{align}\label{eq:defS_TB}
\langle Mv,\phi v\rangle\ge 0, \forall v\in\ell_2,\; M\in\mathcal{M}_{LTV}^{T,B}.
\end{align}

 The links between the sets $\mathcal{S}$ and $\mathcal{S}^{T,B}$ are  explained in the following lemma.
\begin{lemma}\label{le:link}
For any $T,B\in\Z^+$, the set $\mathcal{S}$ is a subset of $\mathcal{S}^{T,B}$. Moreover, it holds
 that
 \begin{align}\label{eq:nonincrease}
     \mathcal{G}_1^{kT,B_1}\subset\mathcal{G}_1^{T,B_2},\;\; \forall k\in\Z^+,B_1\ge B_2
 \end{align}
 and
 \begin{align}\label{eq:convergence}
     \lim_{n\rightarrow\infty}\mathcal{G}_1^{2^n,n}=\mathcal{G}_1:= \{(v,w)\in \ell_2:w=\phi v,\phi\in\mathcal{S}\}.
 \end{align}
\end{lemma}

\noindent\textbf{Proof.} First, we show that $\mathcal{S}\subset\mathcal{S}^{T,B}$ for any $T,B\in\Z^+$. Let $\phi$ be in $\mathcal{S}$. Since $\phi$ is memoryless bounded and monotone, any $(v,w)\in\ell_2$ satisfying $w=\phi v$ is necessarily similarly ordered and unbiased. Thus, for all $v\in\ell_2$, $(v,\phi v)$ satisfies that 
\[
\langle Mv,\phi v \rangle\ge 0,\forall M\in\mathcal{M}_{LTV}.
\]
As  $\mathcal{M}_{LTV}^{T,B}\subset\mathcal{M}_{LTV}$ by definition, $\phi$ is in $\mathcal{S}^{T,B}$.

Next, the result in \eqref{eq:nonincrease} follows directly from the fact that $\mathcal{M}_{LTV}^{kT,B_1}\supset\mathcal{M}_{LTV}^{T,B_2}$ when $k\in\Z^+, B_1\ge B_2$. 

To show \eqref{eq:convergence}, we first prove that 
$\mathcal{G}^{\infty}:=\lim_{n\rightarrow\infty}\mathcal{G}^{2^n,n}$ exists. To this end, we define the sequence $\mathcal{G}^{2^n,n},n=1,2,\ldots$. According to \eqref{eq:nonincrease}, the sequence is monotone non-increasing, and thus the limit exists, i.e., $\mathcal{G}^{\infty}=\cap_{n\in\Z^+}\mathcal{G}^{2^n,n}$. Then, we show in the following that $\mathcal{G}^{\infty}=\mathcal{G}_3$. To see $\mathcal{G}^{\infty}\supset\mathcal{G}_3$, suppose $(v,w)$ is in $\mathcal{G}_3$. Then $(v,w)$ satisfies $\langle Mv,w \rangle \ge  0$ for all $M\in\mathcal{M}_{LTV}$.  It follows that $\langle Mv,w \rangle \ge  0$ holds for all $M\in\mathcal{M}_{LTV}^{T,B}$ and hence $(v,w)\in\mathcal{G}^{T,B}$ for all $T,B\in \Z^+$. Hence $(v,w)\in  \mathcal{G}^{\infty}$. To see $\mathcal{G}^{\infty}\subset\mathcal{G}_3$, suppose by contraposition that $(v,w)\notin \mathcal{G}_3$. Then it follows that there exist $M\in\mathcal{M}_{LTV}$ and $\epsilon>0$ such that $\langle Mv,w\rangle<-\epsilon$. One can always find an $M_1\in\mathcal{M}^{T,B}$ for some large enough $T,B\in\Z^+$ such that $\langle (M_1-M)v,w\rangle<\epsilon/2$. Then 
\begin{align*}
    \langle M_1v,w\rangle=\langle(M_1-M)v,w \rangle+\langle Mv,w\rangle\le -\epsilon/2,
\end{align*}
which implies that $(v,w)\notin \mathcal{G}^{\infty}$. Thus, $\mathcal{G}^{\infty}=\mathcal{G}_3$ is proved. Equation \eqref{eq:convergence} follows immediately by taking the closure of $\mathcal{G}^{\infty}$ and $\mathcal{G}_3$.
\hfill\qed

\vspace{2mm}

Again, it should be noted that Theorem~\ref{th:GTB_pos}  and the preceding lemma  remains  true with $v,w\in \ell_2^{0+}$ since $\ell_2^{0+}\subset\ell_2$.

\vspace{2mm}
\subsection{Robust Stability with LTV Multipliers}
\label{sec:WBmult}

In this subsection, we show that the existence of an appropriate LTV multiplier is necessary and sufficiency for establishing the uniform robust stability of $[G,\phi]$. The sufficiency of the results are formulated using a modern IQC formulation in contrast with the results in~\cite{Willems68}.

\begin{theorem} \label{th: suf_LTV} 
 Let  $G\in \mathcal{L}(\ell_{2e}^{0+}, \ell_{2e}^{0+})$ be LTI, causal and bounded. Let $T,B$ be in $\Z^+$, and
assume the Lurye system $[G,\phi]$ is well-posed for all $\phi \in \mathcal{S}^{T,B}$. The feedback interconnection $[G,\phi]$ is uniformly robustly stable over $\mathcal{S}^{T,B}$ 
if 
\begin{align}\label{eq: sufLTV}
    & \exists M \in\mathcal{M}_{LTV}^{T,B},\; \epsilon>0 \nonumber\\
     & \text{s.t.}\;\; \langle MGw,w \rangle \le -\epsilon \|w\|^2, \forall w\in \ell_2^{0+}.
\end{align}
\end{theorem}
\textbf{Proof:} 
The proof follows from existing results on integral quadratic constraints (IQCs). Specifically,  note  from the definition of $\mathcal{S}^{T,B}$ that if $\phi\in\mathcal{S}^{T,B}$, then $\lambda \phi \in \mathcal{S}^{T,B}$ for all $\lambda\in[0,1]$. Then, it follows from  Theorem \ref{th:GTB_pos} that for every  $\phi\in\mathcal{S}^{T,B}$, it holds that 
\[
\Bigg\langle
\begin{bmatrix}
v \\ \lambda \phi v
\end{bmatrix} ,
\begin{bmatrix}
0 & M^*\\ M &  0
\end{bmatrix}
\begin{bmatrix}
v \\ \lambda \phi v
\end{bmatrix} 
\Bigg\rangle \ge 0
\]
for all $\lambda\in[0,1], v\in\ell_2^{0+}$, and all $M \in \mathcal{M}_{LTV}^{T,B}$. 
By hypothesis there exist  $M\in\mathcal{M}_{LTV}^{T,B}$ and $\epsilon>0$ such that $\langle MGw,w \rangle \le -\epsilon \|w\|^2,\forall w\in \ell_2^{0+}$. Since $G$ is bounded, it implies that there exist $M\in\mathcal{M}_{LTV}^{T,B}$ and $\epsilon>0$ such that
\[
\Bigg\langle
\begin{bmatrix}
Gw \\ w
\end{bmatrix} ,
\begin{bmatrix}
0 & M^*\\ M &  0
\end{bmatrix}
\begin{bmatrix}
Gw \\ w
\end{bmatrix} 
\Bigg\rangle \le - \epsilon \left\| \begin{bmatrix}
Gw \\ w
\end{bmatrix} \right\|^2.
\]
 Thus it follows from the IQC theorem \cite[Corollary IV.3]{Khong21iqc} that the feedback system $[G,\phi]$ is  robustly stable against all $\phi\in\mathcal{S}^{T,B}$. Uniform robust stability follows from the proof in \cite[Theorem 6]{KHONG:21}. 
\hfill\qed

\vspace{2mm}

Before we can develop necessary conditions, we require a version of the lossless S-procedure dealing with LTV quadratic constraint. A time-invariant lossless S-procedure was presented in \cite{Jonsson01}. Here we provide an  S-lossless lemma that involves  time-varying quadratic forms based on the S-procedure lossless theorem in \cite{Jonsson01}.

Recall that  the shift operator $S_\tau$ is defined by $(S_\tau f)_k = f_{k - \tau}$ for $\tau \in \mathbb{Z}_0^+$. Define the quadratic forms $\sigma_k : \ell_2^{0+} \to \Real$ as
\[
\sigma_k(f) = \langle  f,  \Pi_k f \rangle, \qquad k = 0, 1, \ldots,N,
\]
where $\Pi_k : \ell_2 \to \ell_2,k=0, 1,\ldots,N$.

\begin{assumption} \label{as: Pi}
Let $T,T_0\in\mathbb{Z}^+$.  Assume that 
\begin{itemize}
    \item   $\Pi_0$  is bounded, linear, self-adjoint, and $T_0$-periodic;
    
    \item $\Pi_k$ is bounded, linear, self-adjoint, and $T$-periodic;

\end{itemize} 
\end{assumption}

\begin{lemma} \label{le: S_prod}
Suppose the quadratic forms $\sigma_k,k=0,1,\ldots,N$  satisfy  Assumption~\ref{as: Pi}  and that  there exists $f^* \in \ell_2^{0+}$  such that $\sigma_k(f^*) >0$ for $k = 1, \ldots,N$. Then the following are equivalent:
\begin{enumerate} \renewcommand{\theenumi}{\textup{(\roman{enumi})}}\renewcommand{\labelenumi}{\theenumi}
\item \label{item: S1} $\sigma_0(f) \leq 0$ for all $f \in \ell_2^{0+}$ such that $\sigma_k(f) \geq 0$ for all $k = 1, 2, \ldots,N$;

\item \label{item: S2} There exists $\alpha_k\ge 0$, $k=1,\ldots,N$ such that 
  \[
\sigma_0(f) + \sum_{k=1}^N \alpha_k\sigma_k(f) \leq 0,\quad \forall  f \in \ell_2^{0+}.
\]
\end{enumerate}
\end{lemma}
A proof of Lemma~\ref{le: S_prod} is provided in an Appendix.

\vspace{2mm}
Before presenting the main theorem, another  supporting lemma  is stated next.

\begin{lemma}\label{le: rearrangement}
Given a pair of sequences $\{v_{1},v_{2},\ldots,v_{n}\}$,  $\{w_{1},w_{2},\ldots,w_{n}\}$, and suppose  $v_1>v_2>\cdots>v_n>0$ and $w_1>w_2>\cdots>w_n>0$. 
Then   $\sum_{i,j=1}^{n}m_{ij}v_{i}w_{j}>0$ for
all nonzero $M=(m_{ij})\in\mathbb{R}^{n\times n}$ that are doubly hyperdominant.
\end{lemma}
\noindent\textbf{Proof.}
Since  $v_1>v_2>\cdots>v_n$ and $w_1>w_2>\cdots>w_n$, it follows from the rearrangement inequality \cite{hardy1952inequalities} that
\[\sum_{i=1}^{n}v_iw_i > \sum_{i=1}^{n}v_iw_{\pi(i)} 
\]
for all permutation $\pi$ except for $\pi(i)=i, i=1,\ldots,n$.
That means
$\sum_{i,j=1}^{n}[I-P]_{ij}v_iw_j>0$ for all $P\in \{\mathcal{P}_n\setminus I\}$ where $\mathcal{P}_n$ denotes all $n\times n$ permutation matrices.
According to the sufficiency proof of Theorem 3.7 in \cite{willems1970analysis},  given any doubly hyperdominant matrix  $M$ with zero excess, it can be written as $M=\sum_{i=1}^{n!} \beta_i(I-P_i)$, where $P_i\in\mathcal{P}_n$ and $\beta_i\ge 0,k=1,\ldots,n!$. Let $P_1=I$. Note that $M$ being nonzero implies that  there exists at least one $i\in \{2,\ldots,n!\}$ such that $\beta_i>0$. Hence,   $\sum_{i,j=1}^{n}m_{ij}v_{i}w_{j}>0$ for
all $M=(m_{ij})$ that are  nonzero  and doubly hyperdominant with zero excess. 

Now for any given doubly hyperdominant matrix $M\in\mathbb{R}^{n\times n}$, define  $m_{i,n+1}:=-\sum_{j=1}^n m_{ij}$, $m_{n+1,j}:=-\sum_{i=1}^{n}m_{ij}$ for $i,j\le n$, and $m_{n+1,n+1}:=\sum_{i,j=1}^n m_{ij}$.  Since the augmented matrix $M_+:=(m_{ij}),i,j=1,2,\ldots,n+1$ is a doubly hyperdominant with zero excess, by considering the sequence $\{v_{1},v_{2},\ldots,v_{n}, 0\}$  $\{w_{1},w_{2},\ldots,w_{n}, 0\}$, it then follows that 
$\sum_{i,j=1}^{n+1}m_{ij}v_iw_j=\sum_{i,j=1}^{n}m_{ij}v_iw_j>0$. 
\hfill\qed

\begin{theorem} \label{th: nec_LTV} 
 Consider the feedback equation
 \begin{align*}
   Gw+e=v.  
 \end{align*} 
 Let  $G\in \mathcal{L}(\ell_{2e}^{0+}, \ell_{2e}^{0+})$ be LTI, causal and bounded, and let $T,B$ in $\Z^+$ be given with $T\ge 2B+1$.  There exists $\gamma>0$ such that 
 \begin{align}\label{eq:uniform bound}
\sup_{(v,w)\in\mathcal{G}_1^{T,B} \;}
\frac{\| w\|}{\| e\|}\le \gamma 
\end{align}
only if \eqref{eq: sufLTV} is satisfied.  
\end{theorem}

\vspace{2mm}
\textbf{Proof:} 
 Define
\begin{align*}
    \sigma_0(v,w) & := \|w\|^2-\gamma^2 \|v-Gw\|^2\\
    \sigma_k(v,w) & :=\langle C_kv, w \rangle ,k=1,2\ldots,N_{T,B},
\end{align*}
where $(v,w)\in\ell_2^{0+}$, $C_k\in\mathcal{C}^{T,B}$ is defined in  \eqref{eq:finite basis} and  $N_{T,B}$ is the number of nonzero elements in $\mathcal{C}^{T,B}$. 

By defining $e:=v-Gw$, one has that there exists $\gamma>0$ such that \eqref{eq:uniform bound} holds only if  $\sigma_0(v,w)\le 0$ for some $\gamma>0$ and for all $(v,w)\in\ell_2^{0+}$ such that $(v,w)\in\mathcal{G}^{T,B}$, which is equivalent to, by Theorem \ref{th:GTB_pos}, that
\begin{align}\label{eq:condition_slemma}
 \sigma_0(v,w)\le 0, \; \text{ for all } (v,w)\in\ell_2^{0+} \text{ such that } \sigma_k(v,w)\ge 0,k=1,\ldots,N_{T,B}.   
\end{align}

Next,  define $v^*,w^*$ such that  $v_k^*=w_k^*:=\frac{1}{k+1}$, for $k=0,1,\ldots, T-1$ and $v_k^*=w_k^*=0$ otherwise. It is clear that  $(v^*, w^*) \in \ell_2^{0+}$, and we show in the following that
 $\langle C v^*, w^*\rangle>0$ for all nonzero $C\in\mathcal{C}^{T,B}$. 

Let $C$ be any nonzero element in $ \mathcal{C}^{T,B}$, we have that 
\[\langle C v^*,  w^*\rangle =\Bigg\langle \begin{bmatrix}
1\\ \vdots\\\frac{1}{T}
\end{bmatrix}, \tilde{C}  \begin{bmatrix}
1\\ \vdots\\\frac{1}{T}
\end{bmatrix} \Bigg \rangle,\] 
where $\tilde{C}\in \mathbb{R}^{T\times T}$ is the corresponding principal submatrix of $C$. $\tilde{C}$ is nonzero as $C$ is nonzero and $T$-periodic. Since every  principle submatrix of a doubly hyperdominant matrix must be doubly hyperdominant. According to Lemma \ref{le: rearrangement}, one has that 
\[
\Bigg\langle \begin{bmatrix}
1\\ \vdots\\\frac{1}{T}
\end{bmatrix}, \tilde{C}  \begin{bmatrix}
1\\ \vdots\\\frac{1}{T}
\end{bmatrix} \Bigg\rangle>0.
\]
Thus,  $\langle C  v^*, w^*\rangle>0$ for all nonzero $C\in\mathcal{C}^{T,B}$. 

Noting that $\sigma_0$ is time-invariant and $\sigma_k,k=1,\ldots,N_{T,B}$ is $T$-periodic, Assumption \ref{as: Pi} can be easily verified.  By invoking  Lemma \ref{le: S_prod}, \eqref{eq:condition_slemma}   holds if and only if there exist $\alpha_k\ge 0, k=1,\ldots,N_{T,B}$ such that 
\begin{equation}\label{eq:slemma}
 \sigma_0(v,w)+\sum_{k=1}^{N_{T,B}}\alpha_k \sigma_k(v,w)\le 0, \;\forall (v,w)\in \ell_2^{0+}.   
\end{equation}

Now consider the subspace $\{(v,w)\in\ell_2^{0+}:v=Gw\}$, equation \eqref{eq:slemma} implies that 
\[
\sum_{k=1}^{N_{T,B}}\langle \alpha_k C_k Gw,  w\rangle\le -\|w\|^2.
\]

The proof is completed by noting that $\sum_{k=1}^{N_{T,B}} \alpha_k C_k\in \mathcal{M}_{LTV}^{T,B}$.

\hfill\qed
\begin{remark}
It should be remarked that  Theorems \ref{th: suf_LTV} and \ref{th: nec_LTV} hold also for the case  with linear and periodic $G\in \mathcal{L}(\ell_{2e}^{0+},\ell_{2e}^{0+})$. Specifically, for any $T_0\in \mathbb{Z}^+$,   the preceding proofs can be employed directly to show the same results in Theorems \ref{th: suf_LTV} and \ref{th: nec_LTV} but with $T_0$-periodic $G$. This can  be observed by that  the proof of Theorem  \ref{th: suf_LTV} does not  require $G$ to be time-invariant, and  the underlying S-lossless lemma for proving Theorem \ref{th: nec_LTV} allows for periodically time-varying  quadratic form $\sigma_0$ as described in Lemma \ref{le: S_prod}. 
\end{remark}

Next, we consider the nonlinearity set $\mathcal{S}$, which is a subset of $\mathcal{S}^{T,B}$ according to Lemma \ref{le:link}.  As with $\mathcal{S}^{T,B}$, we show in the following corollary that the uniform robust stability of $[G,\phi]$ over $\mathcal{S}$ can be ensured by the existence of a suitable LTV multiplier. 

\begin{corollary}\label{cor:LTV_suf}
 Let  $G\in \mathcal{L}(\ell_{2e}^{0+}, \ell_{2e}^{0+})$ be LTI, causal and bounded. 
Assume the Lurye system $[G,\phi]$ is well-posed for all $\phi \in \mathcal{S}$. The feedback interconnection $[G,\phi]$ is uniformly robustly stable over $\mathcal{S}$ 
if there exist $M\in\mathcal{M}_{LTV}$  and $\epsilon>0$ such that $\langle MGw,w \rangle \le -\epsilon \|w\|^2$ for all $w\in \ell_2^{0+}$.  
\end{corollary}
\noindent\textbf{Proof.} The claim follows from Theorem \ref{th:super} and the same reasoning lines as in the  proof of Theorem \ref{th: suf_LTV}. 
\hfill\qed

\vspace{2mm}

\begin{remark}
Recall from Lemma \ref{le:link} that  $
  \lim_{n\rightarrow\infty}\mathcal{G}_1^{2^n,n}=\mathcal{G}_1:= \{(v,w)\in \ell_2:w=\phi v,\phi\in\mathcal{S}\}.
$ Therefore, if Theorem \ref{th: nec_LTV} could be extended to the case with $(T,B)=\lim_{n\rightarrow\infty}(2^n,n)$, then it could be implied  that $[G,\phi]$ is uniformly robustly stable over $\mathcal{S}$ 
only if  there exist $M\in\mathcal{M}_{LTV}$ and $\epsilon>0$ such that $\langle MGw,w \rangle \le -\epsilon \|w\|^2$ for all $w\in \ell_2^{0+}$.   However,
Theorem \ref{th: nec_LTV} can not be extended to the case with $(T,B)=\lim_{n\rightarrow\infty}(2^n,n)$ since the S-lossless lemma stated in Lemma \ref{le: S_prod} is no longer applicable. In particular, the number of quadratic forms $N_{T,B}$  needed to establish Theorem \ref{th: nec_LTV} will approach infinity as either $T$ or $B$ increases to infinity. By Lemma \ref{le:link}, an infinite number of quadratic forms are required to fully characterise the set $\mathcal{S}$, which hinders the use of the S-lossless lemma. On the other hand, if there were an S-lossless lemma that allows for infinite number of quadratic forms, then the condition in Corollary \ref{cor:LTV_suf} is also necessary whereby the discrete-time Carrasco conjecture can be proved based on the results in Section \ref{sec:OZFmult}. 
\end{remark}

\subsection{Robust Stability with LTI Multipliers}
\label{sec:OZFmult}
By constraining the sets of LTV multipliers previously introduced  to be LTI, we define the following sets of LTI multipliers:
\begin{align*}
    \mathcal{M}_{LTI} &:= \{M\in\mathcal{M}_{LTV}: M \text{ is LTI}\}\\
    \mathcal{M}_{LTI}^{B} &:= \{M\in\mathcal{M}_{LTV}^{T,B}: M \text{ is LTI}\}.
\end{align*}
In fact, $\mathcal{M}_{LTI}^{B}$  is equal, by definition, to $\mathcal{M}_{LTV}^{1,B}$. Every element in $\mathcal{M}_{LTI}^{B}$ represents an LTI system with finite-impulse response. 

In this subsection, we show that LTV multipliers are ``equivalent'' to LTI multipliers~\cite{carrasco2013equivalence}.

\begin{lemma}\label{le:LTV_LTI_TB}
Given any $\epsilon > 0$ and $T,B\in\mathbb{Z}^+$,  there exists $M\in\mathcal{M}_{LTV}^{T,B}$ such that $\langle MGw,w \rangle \le -\epsilon \|w\|^2$ for all $w\in \ell_2^{0+}$ if and only if there exists  $\tilde{M} \in\mathcal{M}_{LTI}^{B}$  such that  $\langle \tilde{M} Gw,w \rangle \le -\epsilon \|w\|^2$ for all $w\in \ell_2^{0+}$.
\end{lemma}
\textbf{Proof:} ($\Leftarrow$)  Sufficiency is obvious as $\mathcal{M}_{LTI}^{B}$ is a subset of $\mathcal{M}_{LTV}^{T,B}$.

($\Rightarrow$) To show necessity, assume  
there exists $M\in\mathcal{M}_{LTV}^{T,B}$ such that $\langle MG w, w \rangle \le -\epsilon \|w\|^2$ for all $w\in \ell_2^{0+}$. 

Note that if $w\in \ell_2^{0+}$ then $S_\tau w\in \ell_2^{0+}$ for all $\tau\in \mathbb{Z}_0^+$. Hence $\langle MGS_\tau w, S_\tau w \rangle \le -\epsilon \|w\|^2$ for all $w\in \ell_2^{0+}$ and $\tau\in \mathbb{Z}_0^+$. It follows from the shift-invariance of $G$ that for all $\tau\in\mathbb{Z}_0^+$ and $w \in \ell_2^{0+}$,
\[
\langle S_{-\tau}MS_\tau  Gw, w \rangle = \langle MS_\tau  Gw, S_{\tau}w \rangle =\langle MGS_\tau w, S_\tau w \rangle \le -\epsilon \|w\|^2 . 
\]
Thus for all $\tau\in\mathbb{Z}_0^+$, the multiplier $S_{-\tau}MS_\tau$ has two useful properties: (i) $S_{-\tau}MS_\tau\in\mathcal{M}_{LTV}^{T,B}$, and (ii) $\langle S_{-\tau}MS_\tau Gw,  w \rangle \le -\epsilon \|w\|^2$ for all $w\in \ell_2^{0+}$. Now define $\tilde{M}:=\frac{1}{T}\sum_{\tau=0}^{T-1}S_{-\tau}MS_\tau$. It follows from (i) that $\tilde{M}\in\mathcal{M}_{LTV}^{T,B}$. Moreover, the fact that  $M$ is $T$-periodic implies  $S_{-1}\tilde{M}S_1=\tilde{M}$. That is, $\tilde{M}$ is LTI. In the end, it follows from (ii) that  $\langle \tilde{M}Gw,w \rangle \le -\epsilon \|w\|^2$ for all $w\in \ell_2^{0+}$. 
\hfill\qed

\vspace{2mm}

The result in Lemma \ref{le:LTV_LTI_TB} can be extended to the case with infinite $T$ and $B$, presented as follows.

\begin{lemma}\label{le:LTV_LTI}
Given any $\epsilon > 0$,  there exists $M\in\mathcal{M}_{LTV}$ such that $\langle MGw,w \rangle \le -\epsilon \|w\|^2$ for all $w\in \ell_2^{0+}$ if and only if there exists  $\tilde{M} \in \mathcal{M}_{LTI}$  such that  $\langle \tilde{M}Gw,w \rangle \le -\epsilon \|w\|^2$ for all $w\in \ell_2^{0+}$.
\end{lemma}

\textbf{Proof.}($\Leftarrow$)  Sufficiency follows from the fact that $\mathcal{M}_{LTI}$ is a subset of $\mathcal{M}_{LTV}$.

($\Rightarrow$) To show necessity, assume 
there exists $M\in\mathcal{M}_{LTV}$ such that $\langle MG w,w \rangle \le -\epsilon \|w\|^2$ for all $w\in \ell_2^{0+}$. 

By the same arguments in the proof of Lemma \ref{le:LTV_LTI_TB}, we have  for all $\tau\in\mathbb{Z}_0^+$, the multiplier $S_{-\tau}MS_\tau$ has three useful properties: (i) $S_{-\tau}MS_\tau\in\mathcal{M}_{LTV}$, (ii) $\langle S_{-\tau}MS_\tau Gw,  w \rangle \le -\epsilon \|w\|^2$ for all $w\in \ell_2^{0+}$, and (iii) $\|S_{-\tau} M S_\tau \| \le \|M\|$.

Next define $M_N\vcentcolon=\frac{1}{N} \sum_{\tau=0}^{N-1} S_{-\tau} M S_\tau$. Boundedness of $S_{-\tau}MS_\tau$ for all $\tau\in \Z_0^+$ implies the sequence $M_N$ is bounded. It follows from Theorems A.3.39 and  A.3.52 in \cite{curtain2012introduction} that there exists a subsequence of $M_N$, denoted as $M_{N_k}$, that is weakly convergent. In other words, there is  $\tilde{M}\in\mathcal{M}_{LTV}$ such that $\lim_{k\to \infty}\langle M_{N_k} v,w \rangle=\langle \tilde{M}v, w\rangle,\;\forall v,w\in\ell_2$.

Define $Y_N \vcentcolon= S_{-1} M_N S_1 - M_N$. Then, it can be observed that $Y_N=\frac{1}{N}(S_{-N} MS_N-M)$.  Since the sequence $S_{-N} M S_{-N}$ is uniformly bounded in the operator norm we have that $Y_N$ converges strongly to zero. 
Hence, we have  
\[
\lim_{N\rightarrow\infty} \langle Y_N  v,  w \rangle =0, \;\forall v,w\in\ell_2.
\]
Considering the subsequence $Y_{N_k}$ of $Y_N$, we further have 
\[
\lim_{k\rightarrow\infty} \big(\langle S_{-1} M_{N_k} S_1  v,  w \rangle -\langle M_{N_k} v,  w \rangle\big) =0, \;\forall v,w\in\ell_2
\]
which leads to
\[
\lim_{k\rightarrow\infty}\langle S_{-1} M_{N_k} S_1 v,   w \rangle = \lim_{k\rightarrow\infty}\langle M_{N_k}  v, w \rangle, \; \forall v,w\in\ell_2.
\]

The right-hand side of the above equation is $\langle \tilde{M} v,  w \rangle$ while the left-hand side can be written as 
$\langle \tilde{M}  S_1v, S_1  w \rangle = \langle S_{-1}\tilde{M}S_1 v, w \rangle$. Thus we have 
\[
\langle S_{-1}\tilde{M}S_1 v, w \rangle = \langle \tilde{M} v,  w \rangle,\; \forall v,w\in\ell_2,
\]
which implies $S_{-1}\tilde{M}S_1=\tilde{M}$ (\cite{young1988introduction}), and therefore $\tilde{M}\in \mathcal{M}_{LTI}$.

Recall that $\langle S_{-\tau} MS_\tau Gw,  w \rangle \le -\epsilon \|w\|^2$ for all $\tau\in\mathbb{Z}_0^+$ and $w\in \ell_2^{0+}$. Therefore $\langle M_N Gw,  w \rangle \le -\epsilon \|w\|^2$ for all $N\in\mathbb{Z}_0^+$ and $w\in \ell_2^{0+}$. 
Since $M_{N_k}$ weakly converges to $\tilde{M}$, it can now be concluded that $\langle \tilde{M} Gw, w\rangle\le -\epsilon \|w\|^2$ for all $w\in \ell_2^{0+}$, which completes the proof.
\hfill \qed

\vspace{2mm}

Based on Theorems \ref{th: suf_LTV} and \ref{th: nec_LTV} and Lemma \ref{le:LTV_LTI_TB}, we present the following theorem.

\begin{theorem}\label{th:iff_LTI}
Let  $G\in \mathcal{L}(\ell_{2e}^{0+}, \ell_{2e}^{0+})$ be LTI, causal and bounded and let $T,B$ in $Z^+$ be given with $T\ge2B+1$. 
\begin{enumerate}
    \item[(i)] \label{item: 1} Assume the Lurye system $[G,\phi]$ is well-posed for all $\phi \in \mathcal{S}^{T,B}$. Then,
the feedback interconnection $[G,\phi]$ is uniformly robustly stable over $\mathcal{S}^{T,B}$  if  
\begin{align}\label{eq: LTI_frequncyCon}
   & \exists M\in \mathcal{M}_{LTI}^{B} \text{ s.t. } \nonumber \\
   & \RE \left\lbrace  \hat{M}(e^{j\omega}) G(e^{j\omega})\right\rbrace < 0,\forall \omega\in [0,2\pi].  
\end{align}
\item[(ii)] \label{item: 2} Consider the feedback equation $Gw+e=v$. Then, \eqref{eq:uniform bound} holds only if  \eqref{eq: LTI_frequncyCon} is satisfied. 
\end{enumerate}

\end{theorem}

\textbf{Proof}: We first show (i). 
Combining  Theorem \ref{th: suf_LTV} and Lemma \ref{le:LTV_LTI_TB}, and assuming well-posedness,
we have that the feedback interconnection  $[G,\phi]$ is uniformly robustly stable over $\mathcal{S}^{T,B}$   if  there exist  
$M \in \mathcal{M}_{LTI}^{B}$  and $\epsilon>0$ such that   $\langle (MG+\epsilon I)w,w \rangle \le 0 $ for all $w\in \ell_2^{0+}$, if and only if there exist  
$M \in \mathcal{M}_{LTI}^{B}$  and $\epsilon>0$ such that   $\big\langle \left((MG+\epsilon I)+(MG+\epsilon I)^*\right)w,w \big\rangle \le 0 $ for all $w\in \ell_2^{0+}$.  By the arguments in  \cite[Theorem 3.1]{MegTre93}, this is then equivalent to $\big\langle \left((MG+\epsilon I)+(MG+\epsilon I)^*\right)w,w \big\rangle \le 0 $ for all $w\in \ell_2$. 
Note that since $M\in\mathcal{M}_{LTI}^{B}$  is bounded and LTI, it admits a transfer function representation $\hat{M}(e^{j\omega}) \in \mathbb{L}_\infty$. Hence, the condition  is  equivalent to that there exist  
$M \in \mathcal{M}_{LTI}^{B}$  and $\epsilon>0$ such that  $\RE\left\lbrace\hat{M}(e^{j\omega}) G(e^{j\omega})+\epsilon I\right\rbrace\le 0$ for all $\omega\in [0,2\pi]$. This is equivalent to there exists  
$M \in \mathcal{M}_{LTI}^{B}$  such that \eqref{eq: LTI_frequncyCon} holds. 

(ii) can be proved analogously by combining Theorem \ref{th: nec_LTV} and Lemma \ref{le:LTV_LTI_TB}.\hfill\qed

\vspace{2mm}

Similarly, combining Corollary \ref{cor:LTV_suf} and Lemma \ref{le:LTV_LTI} leads to the  discrete-time version of the classical Zames-Falb theorem \cite{Zames68} as follows. 
\begin{corollary}
 Let  $G\in \mathcal{L}(\ell_{2e}^{0+}, \ell_{2e}^{0+})$ be LTI, causal and bounded. 
Assume the Lurye system $[G,\phi]$ is well-posed for all $\phi \in \mathcal{S}$.
The feedback interconnection $[G,\phi]$ is uniformly robustly stable over $\mathcal{S}$ if there exists  $M\in \mathcal{M}_{LTI}$  such that $\RE \left\lbrace  \hat{M}(e^{j\omega}) G(e^{j\omega})\right\rbrace < 0$, for all $ \omega\in [0,2\pi]$. 
\end{corollary}

\section{Robust Stability with Nonlinear Multipliers} \label{sec: extention}
In this subsection, we show that the uniform robust stability of $[G,\phi]$ over $\mathcal{S}$ can be also ensured by the existence of an appropriate nonlinear multiplier. This, in turns, provides a possible direction to disprove Carrasco conjecture. 

Recall from Theorem \ref{th:super} that $\mathcal{G}_1$ is equal to the closure of $\mathcal{G}_3$, i.e., the input-output pairs of nonlinearities in $\mathcal{S}$ is can be exactly characterised by the set of multipliers $\mathcal{M}_{LTV}$. Based on this, the sufficiency of the  condition proposed in Corollary \ref{cor:LTV_suf} is established by applying the IQC theorem.  As far as sufficiency is concerned, one may propose a possibly less conservative condition for ensuring the robust stability of the Lurye system by looking for a richer set of multipliers than $\mathcal{M}_{LTV}$ that also fully characterise the  set $\mathcal{S}$. This is shown in what follows. 

Let  $\Psi$ consist of all bounded  memoryless possibly time-varying nonlinear operators $\psi$ that are Lipschitz continuous and satisfy the sector condition $\psi(x,k)x\ge 0,\forall k\in\mathbb{Z},\forall x\in\mathbb{R}$ and $\psi(0,k)=0,\forall k\in\mathbb{Z}$. Let $\mathcal{S}_0$ denote the set of all bounded memoryless monotone nonlinearities that are Lipschitz continuous. 

\begin{lemma}\label{le:nonlinear multiplier}
The set $\mathcal{G}_1  :=\{ (v, w) \in \ell_2 : w = \phi v, \, \phi \in \mathcal{S} \}$ is equal
to the closure of the set
$\mathcal{G}_5 := \{ (v, w) \in \ell_2 :\langle  M\phi_0v, w\rangle + \langle \psi(v,k),w\rangle \ge 0, \forall M\in\mathcal{M}_{LTV}, \forall \phi_0\in\mathcal{S}_0, \forall \psi\in \Psi\}$.
\end{lemma}
\noindent \textbf{Proof.} ($\supset$) Given any $(v,w)\in\mathcal{G}_5$, by letting $\phi_0=I,\psi=0$, it then follows that  $(v,w)\in\mathcal{G}_3$. Thus the claim follows from Theorem \ref{th:super}. 

($\subset$)  Given any $(v,w)\in \mathcal{G}_1$, then $(v,w)$ are similarly ordered and unbiased. Note that for all $\phi_0\in\mathcal{S}_0$, the sequence pairs $(\phi_0 v, w)$ are also similarly ordered and unbiased. Therefore, we have that 
\[\langle  M\phi_0v, w\rangle \ge 0, \;\forall M\in\mathcal{M}_{LTV}.\] 
Moreover,  since  $(v,w)$ is unbiased, we have that   $\psi(v_k,k)w_k\ge 0$ for all $\psi\in\Psi$ whereby 
\[\langle \psi(v,k),w \rangle\ge 0, \; \forall \psi\in\Psi.\] 
Combining the two inequalities above  gives that $(v,w)\in\mathcal{G}_5$, which completes the proof. 
\hfill\qed

\begin{theorem}\label{th: LTV_suf_nonlinear}
 Let  $G\in \mathcal{L}(\ell_{2e}^{0+}, \ell_{2e}^{0+})$ be LTI, causal and bounded. 
Assume the Lurye system $[G,\phi]$ is well-posed for all $\phi \in \mathcal{S}$. The feedback interconnection $[G,\phi]$ is uniformly robustly stable over $\mathcal{S}$
if there exist $M\in\mathcal{M}_{LTV}$, $\psi\in\Psi$, $\phi_0\in\mathcal{S}_0$ and $\epsilon>0$ such that 
\begin{align}\label{eq: condition_nonlinear}
    \langle M\phi_0 G w, w \rangle + \langle \psi(Gw,k),w \rangle \le -\epsilon \|w\|^2,  \forall w\in \ell_2^{0+}.
    \end{align}
\end{theorem}
\noindent\textbf{Proof.} 
Let $\vec{\delta}(\cdot,\cdot)$ denote the directed gap between two systems as defined in \cite{georgiou1997robustness}. A system $\Pi:\ell_2\rightarrow\ell_2$ is said to be \textit{incrementally $L_2$-bounded} if 
\begin{align*}
    \sup_{x,y\in\ell_2: x\neq y} \frac{\|\Pi x- \Pi y\|}{\|x-y\|}< \infty.
\end{align*}

Let $\phi$ be any element in $\mathcal{S}$. In what follows, we show by applying Theorem IV.2 in \cite{Khong21iqc} that $[G,\phi]$ is stable if the condition of Theorem \ref{th: LTV_suf_nonlinear} is satisfied. 

Firstly,  it is clear that $\lambda\in[0,1]\rightarrow \lambda \phi$ is continuous in the directed gap as $\vec{\delta}(\lambda_0\phi, \lambda_1 \phi)\le |\lambda_1-\lambda_0|\|\phi\|$ for all $\lambda_0,\lambda_1\in[0,1]$.
Secondly, note that $[G,0]$ is stable, and that by assumption $[G,\lambda \phi]$ is well-posed for all $\lambda\in[0,1]$. Next, 
from  Lemma \ref{le:nonlinear multiplier} we have that
\[
\Bigg\langle
\begin{bmatrix}
v \\ \phi v
\end{bmatrix} ,
\begin{bmatrix}
0 & 0 \\ M \phi_0 + \psi(\cdot,k )&  0
\end{bmatrix}
\begin{bmatrix}
v \\ \phi v
\end{bmatrix} 
\Bigg\rangle \ge  0
\]
for all $v\in\ell_2^{0+}$, $M \in \mathcal{M}_{LTV}$ and all $\phi_0\in\mathcal{S}_0$, $\psi\in\Psi$. By hypothesis there exist  $M\in\mathcal{M}_{LTV}$, $\phi_0\in\mathcal{S}_0$, $\psi\in\Psi$ and $\epsilon>0$ such that \eqref{eq: condition_nonlinear} holds. Since $G$ is bounded, it follows that there exist $M\in\mathcal{M}_{LTV}$, $\phi_0\in\mathcal{S}_0$,  $\psi\in\Psi$  and $\hat{\epsilon}>0$ such that
\[
\Bigg\langle
\begin{bmatrix}
Gw \\ w
\end{bmatrix} ,
\begin{bmatrix}
0 & 0 \\ M\phi_0+\psi(\cdot, k) &  0
\end{bmatrix}
\begin{bmatrix}
Gw \\ w
\end{bmatrix} 
\Bigg\rangle \le - \hat{\epsilon} \left\| \begin{bmatrix}
Gw \\ w
\end{bmatrix} \right\|^2.
\]
Since $M\in\mathcal{M}_{LTV}$ is linear and $\phi_0\in\mathcal{S}_0$ is memoryless and Lipschitz continuous,  one can show $M\phi_0$ is incrementally $\ell_2$-bounded. Similarly, $\psi\in\Psi$ being memoryless and Lipschitz leads to that $\psi$ is also incrementally $\ell_2$-bounded. Hence, $M\phi_0+\psi(\cdot, k)$ is incrementally $\ell_2$-bounded. Now, by invoking Theorem IV.2 in \cite{Khong21iqc} we obtain that $[G,\lambda \phi]$ is uniformly stable over $\lambda\in[0,1]$. Since $\phi$ can be any element in $\mathcal{S}$,  it follows that $[G,\phi]$ is robustly stable for all $\phi\in\mathcal{S}$. 
Uniform robust stability follows from the proof in \cite[Theorem 6]{KHONG:21}. 
\hfill\qed
\begin{remark}
The sufficient condition in Corollary \ref{cor:LTV_suf} is at least as conservative as the condition in Theorem \ref{th: LTV_suf_nonlinear}.
This can be seen from that the condition in Corollary \ref{cor:LTV_suf} can be recovered from the condition in Theorem \ref{th: LTV_suf_nonlinear} by fixing $\phi_0=I$ and $\psi=0$. It is noteworthy that this is the first time that a nonlinear multiplier is proposed to establish feedback stability of the Lurye system with monotone nonlinearities. 
\end{remark}

 Lemma  \ref{le:nonlinear multiplier} shows that the set $\mathcal{S}$ can be equivalently characterised also by the set of  multipliers that involves both LTV multiplier in $\mathcal{M}_{LTV}$ and nonlinear multipliers in  $\Psi$ and $\mathcal{S}_0$. It is  shown  in the preceding subsection  that the existence of $M\in\mathcal{M}_{LTV}$ is ``equivalent" to the existence of $\mathcal{M}_{LTI}$, but this is unlikely to  hold for the nonlinear multipliers. In fact, the results in this work provide a possible direction to disprove Carrasco conjecture. That is  to find a counterexample $G$ such that there is no $M\in \mathcal{M}_{LTI}$ satisfying \eqref{eq: LTI_frequncyCon} but there exist $M\in \mathcal{M}_{LTV}, \phi_0\in\mathcal{S}_0, \psi\in\Psi$ and $\epsilon>0$ satisfying \eqref{eq: condition_nonlinear}.  

\section{Conclusion}
Motivated by the discrete-time Carrasco conjecture, this work studied both the  necessity and sufficiency of a suitable LTI multiplier for uniform robust stability of discrete-time Lurye systems.
First,  it is  shown that the set of monotone nonlinearities is fully characterised by a set of LTV multipliers.  Significantly, we show that a conic parameterization of LTV multipliers is possible. By restricting the set of LTV multipliers to be banded and periodic,  we introduced a larger set of nonlinearities.  Second, it is shown that the existence of a suitable banded and periodic LTV multiplier is sufficient for establishing the uniform robust stability of the Lurye system with time-varying periodic plant over the larger  set of nonlinearities. The same condition is also shown  to be necessary when the nonlinearity set is replaced by the relation set that is characterised by the same set of LTV multipliers. Third, when the pant is LTI, the existence of such a suitable LTV multiplier is shown to be equivalent to the existence of a suitable LTI multipliers. 

The sufficiency direction in the second step above can be extended straightforwardly to the case that considers the set of monotone nonlinearities. This enables the recovery of the discrete-time Zames-Falb theorem. However, it remains unknown that if the necessity direction can be extended similarly. This leaves us an interesting future direction. If  this can be done, then one can prove the converse of the discrete-time Zames-Falb theorem, and thus prove the discrete-time Carrasco conjecture. 
Another possible  direction is to disprove the discrete-time Carrasco conjecture by searching for a counterexample such that a suitable nonlinear multiplier exists while no suitable LTI multipliers exist. 

In continuous-time, the Carrasco conjecture remains open, and each of the three required steps poses interesting research challenges.

\bibliographystyle{plain} 
\bibliography{zf_reference}

\appendix
\section{Proof of Lemma~\ref{le: S_prod}}
%\noindent\textbf{Proof.}
 That \ref{item: S2} implies \ref{item: S1} is obvious.  To see that \ref{item: S1} implies \ref{item: S2}, define
  \begin{align*}
  \mathcal{K} & = \left\{\left(\sigma_0(f), \sigma_1(f), \ldots, \sigma_N(f)  \right) : f\in \ell_2^{0+}\right\} \\
    \mathcal{N} & = \left\{(n_0, n_1, \ldots,n_N) :  n_k > 0 \; \forall k \in \{0,1,\ldots,N\}\right\}.
  \end{align*}
  We show below that $\overline{\mathcal{K}}$, the closure of $\mathcal{K}$, is convex. Let
  $f_1, f_2 \in \ell_2^{0+}$, 
  \begin{align*}
    k_1 &= \left(\sigma_0(f_1), \sigma_1(f_1), \ldots, \sigma_N(f_1)\right) \in \mathcal{K}  \\
    k_2 &= \left(\sigma_0(f_2), \sigma_1(f_2), \ldots, \sigma_N(f_2)\right) \in \mathcal{K}.
  \end{align*}
  For all $\lambda \in [0, 1]$,
  \begin{align*}
    & \sigma_k(\sqrt{\lambda} f_1 + \sqrt{1 - \lambda} S_\tau f_2) \\
    = \; & \lambda \sigma_k(f_1) + (1 - \lambda) \sigma_k(S_\tau f_2) + 2\sqrt{\lambda(1-\lambda)} \langle \Pi_k f_1,  S_\tau f_2 \rangle.
  \end{align*}

Observe that $ \langle \Pi_k f_1, S_\tau f_2 \rangle  \to 0, k=0,1,\ldots N$, as $\tau \to \infty$.  Moreover, since $\Pi_k,k=1,\ldots N$ is $T$-periodic and $\Pi_0$ is $T_0$-periodic, it follows that for all $\epsilon > 0$ there exists sufficiently large $\beta \in \mathbb{Z}^+$ such that with $\tau := \beta TT_0$,
\begin{align*}
  \Big| \langle \Pi_k  f_1, S_\tau f_2 \rangle \Big| & < \epsilon,\; \forall k=0,1,\ldots,N \quad \text{and} \\
    \sigma_k(S_\tau f_2) - \sigma_k(f_2)  & = 0, \; \forall k=0,1,\ldots,N.
  \end{align*}
Together, we have that 
\[
\left( \sigma_0(\sqrt{\lambda} f_1 + \sqrt{1 - \lambda} S_\tau f_2), \ldots,\sigma_N(\sqrt{\lambda} f_1 + \sqrt{1 - \lambda} S_\tau f_2) \right)\rightarrow\lambda k_1 + (1 - \lambda)k_2
\]
as $\beta\rightarrow\infty$. That is, $\overline{\mathcal{K}}$ is convex.

Since $\mathcal{N}$ is open and \ref{item: S1} implies that $\overline{\mathcal{K}} \cap \mathcal{N} = \emptyset$, it follows from the  hyperplane separation theorem that there exists a  hyperplane that separates $\overline{\mathcal{K}}$ and $\mathcal{N}$. That is, there exists a nonzero $N+1$-tuple $(c_0,c_1,\ldots,c_N)$ such that 
\begin{align} \label{eq: sep_H}
c_0n_0+c_1n_1+\cdots+c_Nn_N > 0 \quad \forall (n_0,n_1,\ldots,n_N) \in \mathcal{N}
\end{align}
and 
\begin{align} \label{eq: sep_K}
c_0\kappa_0+c_1\kappa_1+\cdots+c_N\kappa_N \leq 0 \quad \forall (\kappa_0,\kappa_1,\ldots,\kappa_N) \in \overline{\mathcal{K}}.
\end{align}
Note that
\eqref{eq: sep_H} being true for all $(n_0,n_1,\ldots,n_N) \in \mathcal{N}$ implies that $c_k\ge 0,k=1,\ldots,N$.

Finally, let $\kappa_k = \sigma_k(f^*)$ for $k =  0,1, \ldots,N$. Note that $\kappa_k  > 0$ for $k = 1, \ldots,N$ by hypothesis. It follows from
\eqref{eq: sep_K} that $c_0>0$. 
Dividing \eqref{eq: sep_K} by $c_0$ and letting  $\alpha_k=c_k/c_0$, $k=1,\ldots,N$ then yields \ref{item: S2}, as required. $\hfill$\qed

\end{document}